\RequirePackage{fix-cm}
\documentclass[twocolumn,epjc3]{svjour3}  
\RequirePackage{graphicx}
\RequirePackage{amsmath,amssymb}
\RequirePackage{mathptmx}
\RequirePackage{isotope}
\RequirePackage{cite}
\usepackage{upgreek}
\usepackage{color}
\usepackage{doi}
\RequirePackage{hyperref}
\RequirePackage{booktabs}
\RequirePackage{siunitx}
\usepackage{enumitem}

\usepackage{tabularx}
\usepackage{bm}

\usepackage{threeparttable}

\RequirePackage{lineno}\newdimen\linenumbersep\linenumbersep=1.5pt\renewcommand{\linenumberfont}\tiny
\journalname{Eur. Phys. J. C}
\begin{document}
\title{Suppression of Penning discharges between the KATRIN spectrometers}
% autogenerated by authorlist.py from input file 'overleaf_author_list.tex' using svjour3 format

% Affiliations:
\institute{%
Institute for Technical Physics~(ITEP), Karlsruhe Institute of Technology~(KIT), Hermann-von-Helmholtz-Platz 1, 76344 Eggenstein-Leopoldshafen, Germany\label{a}
\and Institute for Nuclear Physics~(IKP), Karlsruhe Institute of Technology~(KIT), Hermann-von-Helmholtz-Platz 1, 76344 Eggenstein-Leopoldshafen, Germany\label{b}
\and Technische Universit\"{a}t M\"{u}nchen, James-Franck-Str. 1, 85748 Garching, Germany\label{c}
\and IRFU (DPhP \& APC), CEA, Universit\'{e} Paris-Saclay, 91191 Gif-sur-Yvette, France \label{d}
\and Institute for Data Processing and Electronics~(IPE), Karlsruhe Institute of Technology~(KIT), Hermann-von-Helmholtz-Platz 1, 76344 Eggenstein-Leopoldshafen, Germany\label{e}
\and Institute of Experimental Particle Physics~(ETP), Karlsruhe Institute of Technology~(KIT), Wolfgang-Gaede-Str. 1, 76131 Karlsruhe, Germany\label{f}
\and Institute for Nuclear Research of Russian Academy of Sciences, 60th October Anniversary Prospect 7a, 117312 Moscow, Russia\label{g}
\and Max-Planck-Institut f\"{u}r Kernphysik, Saupfercheckweg 1, 69117 Heidelberg, Germany\label{h}
\and Max-Planck-Institut f\"{u}r Physik, F\"{o}hringer Ring 6, 80805 M\"{u}nchen, Germany\label{i}
\and Department of Physics and Astronomy, University of North Carolina, Chapel Hill, NC 27599, USA\label{j}
\and Triangle Universities Nuclear Laboratory, Durham, NC 27708, USA\label{k}
\and Department of Physics, Faculty of Mathematics and Natural Sciences, University of Wuppertal, Gau\ss{}str. 20, 42119 Wuppertal, Germany\label{l}
\and Departamento de Qu\'{i}mica F\'{i}sica Aplicada, Universidad Autonoma de Madrid, Campus de Cantoblanco, 28049 Madrid, Spain\label{m}
\and Center for Experimental Nuclear Physics and Astrophysics, and Dept.~of Physics, University of Washington, Seattle, WA 98195, USA\label{n}
\and Nuclear Physics Institute of the CAS, v. v. i., CZ-250 68 \v{R}e\v{z}, Czech Republic\label{o}
\and Institut f\"{u}r Kernphysik, Westf\"{a}lische Wilhelms-Universit\"{a}t M\"{u}nster, Wilhelm-Klemm-Str. 9, 48149 M\"{u}nster, Germany\label{p}
\and Helmholtz-Institut f\"{u}r Strahlen- und Kernphysik, Rheinische Friedrich-Wilhelms-Universit\"{a}t Bonn, Nussallee 14-16, 53115 Bonn, Germany\label{q}
\and Laboratory for Nuclear Science, Massachusetts Institute of Technology, 77 Massachusetts Ave, Cambridge, MA 02139, USA\label{r}
\and Department of Physics, Carnegie Mellon University, Pittsburgh, PA 15213, USA\label{s}
\and Institute for Nuclear and Particle Astrophysics and Nuclear Science Division, Lawrence Berkeley National Laboratory, Berkeley, CA 94720, USA\label{t}
\and University of Applied Sciences~(HFD)~Fulda, Leipziger Str.~123, 36037 Fulda, Germany\label{u}
\and Department of Physics, Case Western Reserve University, Cleveland, OH 44106, USA\label{v}
\and Institut f\"{u}r Physik, Johannes-Gutenberg-Universit\"{a}t Mainz, 55099 Mainz, Germany\label{w}
\and Institut f\"{u}r Physik, Humboldt-Universit\"{a}t zu Berlin, Newtonstr. 15, 12489 Berlin, Germany\label{x}
\and Project, Process, and Quality Management~(PPQ), Karlsruhe Institute of Technology~(KIT), Hermann-von-Helmholtz-Platz 1, 76344 Eggenstein-Leopoldshafen, Germany    \label{y}
}

% Authors:
\author{%
M.~Aker\thanksref{a,b}
\and K.~Altenm\"{u}ller\thanksref{c,d}
\and A.~Beglarian\thanksref{e}
\and J.~Behrens\thanksref{b,f}
\and A.~Berlev\thanksref{g}
\and U.~Besserer\thanksref{a,b}
\and K.~Blaum\thanksref{h}
\and F.~Block\thanksref{f}
\and S.~Bobien\thanksref{a}
\and B.~Bornschein\thanksref{a,b}
% ...
}

\maketitle
\newpage
\begin{abstract}

The KArlsruhe TRItium Neutrino experiment (KATRIN) aims to determine the effective electron (anti)-neutrino mass with a sensitivity of \SI{0.2}{eV/c^2} by precisely measuring the endpoint region of the tritium $\beta$-decay spectrum. It uses a tandem of electrostatic spectrometers working as MAC-E (magnetic adiabatic collimation combined with an electrostatic) filters. In the space between the pre-spectrometer and the main spectrometer, creating a Penning trap is unavoidable when the superconducting magnet between the two spectrometers, biased at their respective nominal potentials, is energized. The electrons accumulated in this trap can lead to discharges, which create additional background electrons and endanger the spectrometer and detector section downstream. To counteract this problem, ``electron catchers" were installed in the beamline inside the magnet bore between the two spectrometers. These catchers can be moved across the magnetic-flux tube and intercept on a sub-ms time scale the stored electrons along their magnetron motion paths. In this paper, we report on the design and the successful commissioning of the electron catchers and present results on their efficiency in reducing the experimental background.

\end{abstract}


\section{Introduction}

The KArlsruhe TRItium Neutrino experiment (KATRIN) \cite{KATRIN_design} at the Karlsruhe Institute of Technology is aiming to determine the average electron (anti)neutrino mass with a sensitivity of \SI{0.2}{eV/c^2}  (90$\%$ C.L.)  \cite{KATRIN} in a direct, model-independent way using a precision measurement of the tritium $\beta$-decay spectrum near the endpoint. The observable in this case is an incoherent sum over the mass eigenstates contributing to the electron (anti)neutrino $\nu_e$, given by
\begin{equation}
	m^2(\nu_e) = \sum_i \vert U_{ei} \vert ^2 m^2(\nu_i),
\end{equation}
where $U_{ei}$ are the elements of the neutrino-mixing matrix connecting the electron neutrino flavour to the $i$th neutrino mass state of mass $m_i$. The former upper limit of $m(\nu_e) \lesssim $ \SI{2}{eV/c^2} on direct mass measurements have been set by the Mainz \cite{Mainz} and Troitsk \cite{Troitsk1, Troitsk2} which has been recently improved by the KATRIN experiment by almost a factor of two: $m(\nu_e) < $ \SI{1.1}{eV/c^2} \cite{KATRIN_result}

\begin{figure*}[ht]
\center
\includegraphics[width=1.\textwidth]{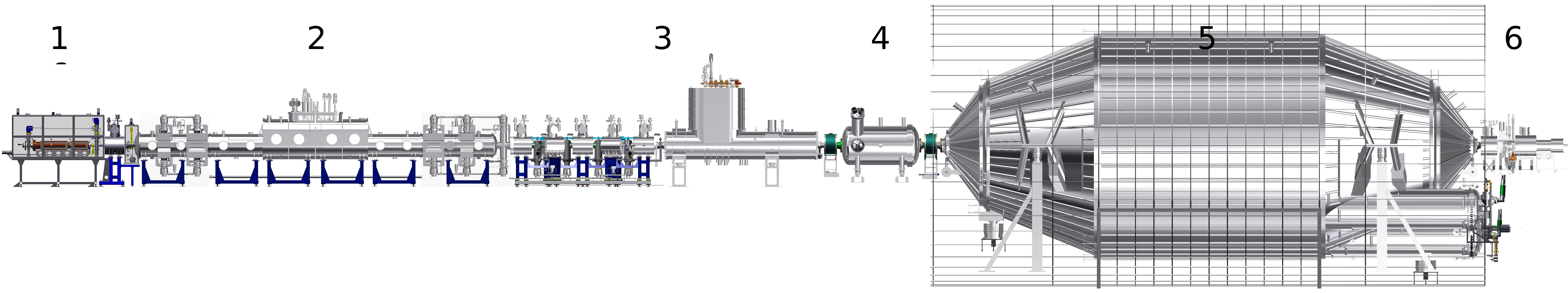}
\caption{Experimental setup of the KATRIN experiment. The main components are: (1) calibration and monitoring rear section, (2) windowless gaseous tritium source, (3) transport section, (4) pre-spectrometer, (5) main spectrometer, and (6) focal plane detector.}
\label{KATRIN_beamline}
\end{figure*}

Figure \ref{KATRIN_beamline} shows an overview of the KATRIN setup. Electrons emitted in the Windowless Gaseous Tritium Source (WGTS) are adiabatically guided by superconducting magnets through the transport section, where tritium is removed by differential and cryogenic pumping. The electrons then enter a tandem of electrostatic spectrometers, known as the pre- and main spectrometers, which operate in MAC-E-filter mode \cite{KATRIN_setup}. The electrons transmitted through both spectrometers are then counted by the focal plane detection system (FPD) \cite{FPD}. 

The key components of a MAC-E spectrometer are a magnetic guiding field and an electrostatic barrier; in KATRIN, the first is created by superconducting solenoids on both ends of each spectrometer. In the center of the main spectrometer, the magnetic field can be fine-tuned by a low-field correction system (LFCS) consisting of vertical air coils surrounding the main spectrometer vessel, and by additional horizontal coils to compensate the Earth magnetic field (EMCS) \cite{air_coil}. $\beta$-electrons enter a spectrometer from the source side and follow the magnetic field lines in cyclotron motion into the spectrometer. On their way, the magnetic field drops by several orders of magnitude, resulting in an almost complete conversion of the cyclotron motion of the adiabatically moving electrons into longitudinal motion. In the middle of the spectrometer, the electrons encounter the maximum of the electrostatic potential at the analyzing plane, and those which have enough energy to overcome it are reaccelerated toward the exit of the spectrometer. In the case of the pre-spectrometer, the electrons then proceed to the main spectrometer, while in the case of the main spectrometer they are counted by the FPD \cite{FPD}, a monolithic, silicon PIN diode with 148 pixels arranged in rings. FPD data are collected by the data-acquisition hardware, which is controlled and read out by ORCA (Object-oriented Real-time Control and Acquisition) software \cite{ORCA}.

The task of the pre-spectrometer is to reflect all low-energy electrons which do not carry information about the neutrino mass. Therefore its design retarding potential is set about \SI{300}{V} below the endpoint of the $\beta$-spectrum, i.e. to \SI{-18.3}{kV}. The main spectrometer analyzes the kinetic energy of the remaining $\beta$-electrons with eV resolution. Its retarding potential is scanned around the endpoint energy within a range between about \SI{-18.5}{kV} and \SI{-18.6}{kV}.

\section{Inter-spectrometer Penning trap problem and countermeasures}
\label{description}

To reach the proposed sensitivity in KATRIN,  the experimental background must be kept low, with a total rate of \SI{10}{mcps}. $\beta$-electrons inside the main spectrometer can produce secondary electrons via a chain of processes, several of which will be discussed below. These low-energy secondary electrons can then be accelerated toward the detector by the electric potential of the main spectrometer to energies around \SI{18.6}{keV}.  At these energies the secondary electrons cannot be distinguished from signal electrons in the endpoint region, so an overall increase in the background level results. To counteract this effect, the pre-spectrometer is used to reduce the flux of the $\beta$-electrons entering the main spectrometer. However, the combination of the retarding potentials (the pre-spectrometer at $U_{\mathrm{PS}} = $\SI{-18.3}{kV}, the main spectrometer at  $U_{\mathrm{MS}} \approx $\SI{-18.6}{kV} and the beamline between them at ground potential) and a magnetic field of up to \SI{4.5}{T} produced by their common superconducting magnet forms a Penning trap for negatively charged particles (see fig. \ref{PT_fields}). An electron passing through this region can lose energy due to processes of elastic and inelastic scattering on residual gas molecules and by cyclotron radiation, thereby becoming trapped. Apart from the $\beta$-electrons from the tritium source, the trap is fed by background from both spectrometers.

Several physical processes are connected to the presence of stored electrons in the inter-spectrometer region: 

\begin{figure}[ht]
\center
\includegraphics[width=\columnwidth]{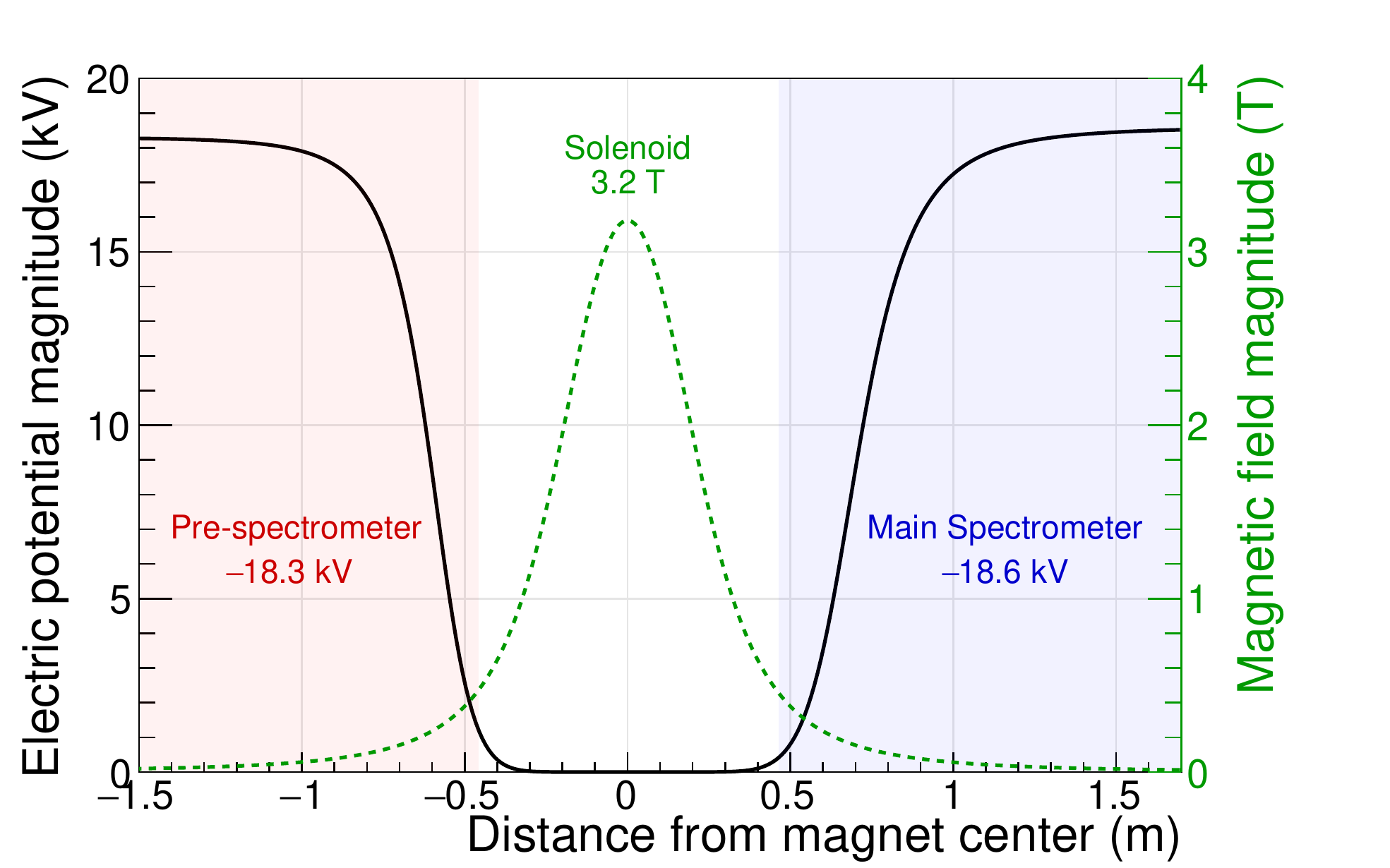}
\caption{Distribution of electric and magnetic fields in the region of the inter-spectrometer Penning trap with $71\%$ of the maximum magnetic field of the solenoid. The shaded areas mark different subsystems: pre-spectrometer (red), inter-spectrometer region (white) and main spectrometer (green).
%The z-axis represent the direction along the beamline.
}
\label{PT_fields}
\end{figure}

\begin{enumerate}[label=(\alph*)]
\item Electrons stored in the Penning trap lose transverse kinetic energy due to cyclotron radiation. The energy loss

 \begin{equation} 
	E_\perp (t) = E_{\perp 0} \cdot e^{-\Gamma t}
\end{equation}
 
($E_\perp$ is the transversal kinetic energy of the electron dependent on time $t$ with a starting value of $E_{\perp 0}$) is characterized by the attenuation factor \cite{KATRIN_design}

\begin{equation} \Gamma =  \frac{1}{\tau_{\mathrm{Cy}}} = \frac{\dot{E}_\perp}{E_\perp} \approx 0.4\textrm{ s}^{-1}\left(\frac{B}{1{\mathrm{T}}}\right)^2,
\end{equation} 
where $B$ is the magnetic field at the electron trajectory and $\tau_{\mathrm{Cy}}$ is the cyclotron cooling time.

The exact losses depend on the magnetic field (which is not constant throughout the trap volume) and on the electron's polar angle to the magnetic field lines $\theta$. Also, cooling via cyclotron radiation can be enhanced by elastic scattering of the electrons on the residual gas molecules, as part of the kinetic energy can be reshuffled into the transverse component which then allows the energy to be radiated away. 

\item Due to inelastic scattering with residual gas molecules, additional electrons are created and trapped. 
There is a significant chance for a trapped primary electron to induce ionization in the high-electric-potential region at either end of the trap shortly before its return point (where it gets reflected by the electric potential); the maximum of the ionization cross section $\sigma^{\mathrm{ion}}_{e}(E)\approx\SI{e-20}{m^2}$ of $\mathrm{H_2}$\footnote{In the ultra-high vacuum conditions of the KATRIN spectrometers, most of the residual gas consists of $\mathrm{H_2}$ molecules.} by electron impact corresponds to electron kinetic energies around \SI{70}{eV} \cite{el_ion}. In this case, a secondary electron, which is trapped as well, can gain enough energy to trigger further ionizations.
The mean time between ionizations at a pressure of $p=\SI{4e-11}{mbar}$ (or the corresponding particle density of  $n({\mathrm{H_2}})=$\SI{9.7e11}{m^{-3}}) and a temperature of \SI{300}{K} is estimated for an electron of velocity $v$ using the maximum cross section at $E=\SI{70}{eV}$ as

\begin{equation} 
\tau_\mathrm{ion} = \frac{1}{\dot{N}_{e}^\mathrm{ion}} = \frac{1}{n v \sigma_{e}^\mathrm{ion}} = \sqrt{\frac{m_e}{2E}}\frac{1}{ n \sigma_{e}^\mathrm{ion}} \approx \SI{20}{s},
\end{equation}

where $\dot{N}_{e}^{\mathrm{ion}}$ is the rate of ionization collisions for a single electron and $m_e$ is the mass of the electron.
This can lead to an avalanche process, creating more and more charged particles.
According to detailed electron tracking calculations, which include electron inelastic and elastic differential scatterings with $\mathrm{H_2}$ molecules (including angular changes) and cyclotron radiation in a \SI{4.5}{T} magnetic field, an initially stored electron with $E_0=\SI{18}{keV}$ starting energy produces $N_{e-\mathrm{i}} = 10^8$ ionizations. This multiplicity is very sensitive to the electron energy: with \SI{10}{keV}, \SI{5}{keV}, and \SI{1}{keV}, it is only 40000, 1000, and 20, respectively. With \SI{1e-11}{mbar} residual gas pressure, the electron cooling time is of the order of \SI{1}{h}.
A large number of electrons can build up a space charge that will even affect signal electrons passing through the Penning trap region.

Due to the good vacuum, the stored electrons cannot leave the Penning trap in radial direction (perpendicular to magnetic field) by collisions with residual gas molecules. Nevertheless, the long duration of the storage of electrons in the Penning trap can be influenced by plasma instabilities, which can exist in the non-neutral (electron) plasma of a Penning discharge. Fluctuating azimuthal electric fields can induce a radial $E\times B$ drift-diffusion of the electrons, which is called anomalous diffusion, and reduces the electron storage time in the trap \cite{crossed-field_instab, el_cloud_instab, rot_struct_in_plasma, fluid_sim_instab}. There are various plasma instabilities; one example, which could be interesting for our Penning trap, is the  cross-field ($E\times B$) instability which requires a radial plasma density gradient and an inwardly directed radial electric field in the trap \cite{nonlin_growth_instab, part-in-cell_sim}.

\item The ions, created along with the secondary electrons, are not trapped inside the inter-spectrometer Penning trap and can leave toward the spectrometers. There, in the low magnetic field, they move non-adiabatically until they hit the spectrometer vessel, having on their way a minor (but still non-zero) probability to produce background electrons due to scattering with residual gas. The ionization collision cross section of $\mathrm{H^+}$ and $\mathrm{H^+_2}$ of 18-keV kinetic energy with $\mathrm{H_2}$ molecules is about $\sigma_{\mathrm{ion}}=\SI{e-20}{m^2}$, which for the pressure of $p=\SI{4e-11}{mbar}$ and a mean ion path length of $l=\SI{24}{m}$ gives an ionization probability of $P_{\mathrm{ion}} = \sigma_{\mathrm{ion}} \cdot n \cdot l = 2.3 \times 10^{-7}$. This means that for each trapped primary electron with a starting energy of \SI{18}{keV}, up to $N_\mathrm{trap}^\mathrm{ionization} \approx N_{e-\mathrm{i}} \cdot P_{\mathrm{ion}} = 23$ electrons can reach the detector.

Apart from ionization of the residual gas, the ions can produce electrons by hitting the metallic surface of the spectrometer walls.  The surface, however, is efficiently shielded by the approximately axisymmetric magnetic field and by the wire electrode system of the spectrometer. 
The lower limit for the suppression factor of the shielding is about $10^{4}$~\cite{muon_KATRIN}.
Assuming again that $10^8$ electron-ion pairs are generated for each primary trapped electron and further assuming that each ion knocks out 10 electrons at the surface, there could be up to $10^5$ electrons reaching the detector per primary trapped electron by this process. 

\item UV photons can be created from the de-excitation of residual gas molecules after colliding with trapped electrons. These photons, not being affected by electric and magnetic fields, can hit metal surfaces and release more electrons, or, in the extreme ultraviolet case, create secondary electrons by photoionization of residual gas molecules, potentially leading to background rates comparable to those caused by ions. 

\item Another recently identified source of background electrons, which was overlooked in the past, is highly excited neutral atoms or molecules, e.g. Rydberg atoms or molecules. Such particles are produced via ion-induced sputtering and can propagate freely within the spectrometers. They can be ionized by infrared photons from the blackbody radiation of the spectrometer vessel (or by auto-ionization if more than one electron of the atom is excited). Electrons from this process feed the trap, and Penning-trap induced ions entering the main spectrometer produce electrons, which cannot be distinguished from normal signal electrons, via the sputtering-Rydberg process. \footnote{It should be noted that in \cite{Prall} the background due to $\beta$-electrons when operating the KATRIN experiment with zero pre-spectrometer potential (as an option to prevent creation of the trap) was calculated to be a negligible value of about \SI{0.001}{mcps}-\SI{0.01}{mcps} at \SI{1e-11}{mbar}, however, the Rydberg-states-related process was not yet considered there.}

\end{enumerate}

In consequence, electrons accumulating in the Penning trap between the pre- and main spectrometers can lead to elevated background rates much larger than tolerable at the KATRIN experiment. An exponentially growing avalanche or discharge may present a danger for the spectrometer and detector section of KATRIN. Not only is the former effect strongly dependent on pressure (as shown in the discussion above), but the latter is as well: the formative time $t_f$ of a Penning discharge is inversely proportional to the pressure $p$ of the residual gas inside the Penning trap: $t_f \propto 1/p$ \cite{form_time}. Therefore, the residual pressure in the spectrometers, along with their electric potentials, is a crucial parameter for the inter-spectrometer Penning trap problem.

Since this Penning trap cannot be avoided in a tandem MAC-E-filter setup and since the filling of the trap by electrons is very difficult to completely avoid, a method to eject stored electrons from the Penning trap before they can ionize residual gas molecules is required. Several ejection processes have already been investigated in the past with a goal of achieving an ejection time shorter than the ionization time:

\begin{equation}
  \tau_\mathrm{eject} < \tau_\mathrm{ion}.
\end{equation}

In an early study it was shown from simulations \cite{dipl_essig} that electrons stored in the Penning trap between the two KATRIN spectrometers will indeed lead to a cascade of secondary electrons producing a significant amount of background. It was also investigated how an $E \times B$ drift could eject stored electrons when applying a transversal electric field, but this method would require very high electric fields which would probably lead to discharge problems. Therefore a mechanical ejection method was developed and tested \cite{beck, dr_valerius} using the spectrometer from the former Mainz Neutrino Mass Experiment \cite{Mainz}. It was shown that a wire sweeping through the Penning trap removed stored electrons and successfully  stopped the build-up of background processes. The position of the wire was controlled by an electric current through the wire, which subsequently moved in the magnetic field due to the Lorentz force. The slowly sweeping wire was able to catch all trapped but fast-moving electrons because the magnetron motion of the trapped electrons caused them to collide with the wire within a very short time (see Section \ref{WorkingPrinciple} below). 
The fact that the magnetron drift is really sufficient to safely remove electrons stored in the Penning trap even under the conditions at KATRIN was investigated in detail in a test experiment at the KATRIN pre-spectrometer \cite{dr_hillen}. Here, the principle of the electron catcher in the form of a thin pin was successfully demonstrated for the first time. 

\begin{figure}[ht]
\center
\includegraphics[width=\columnwidth]{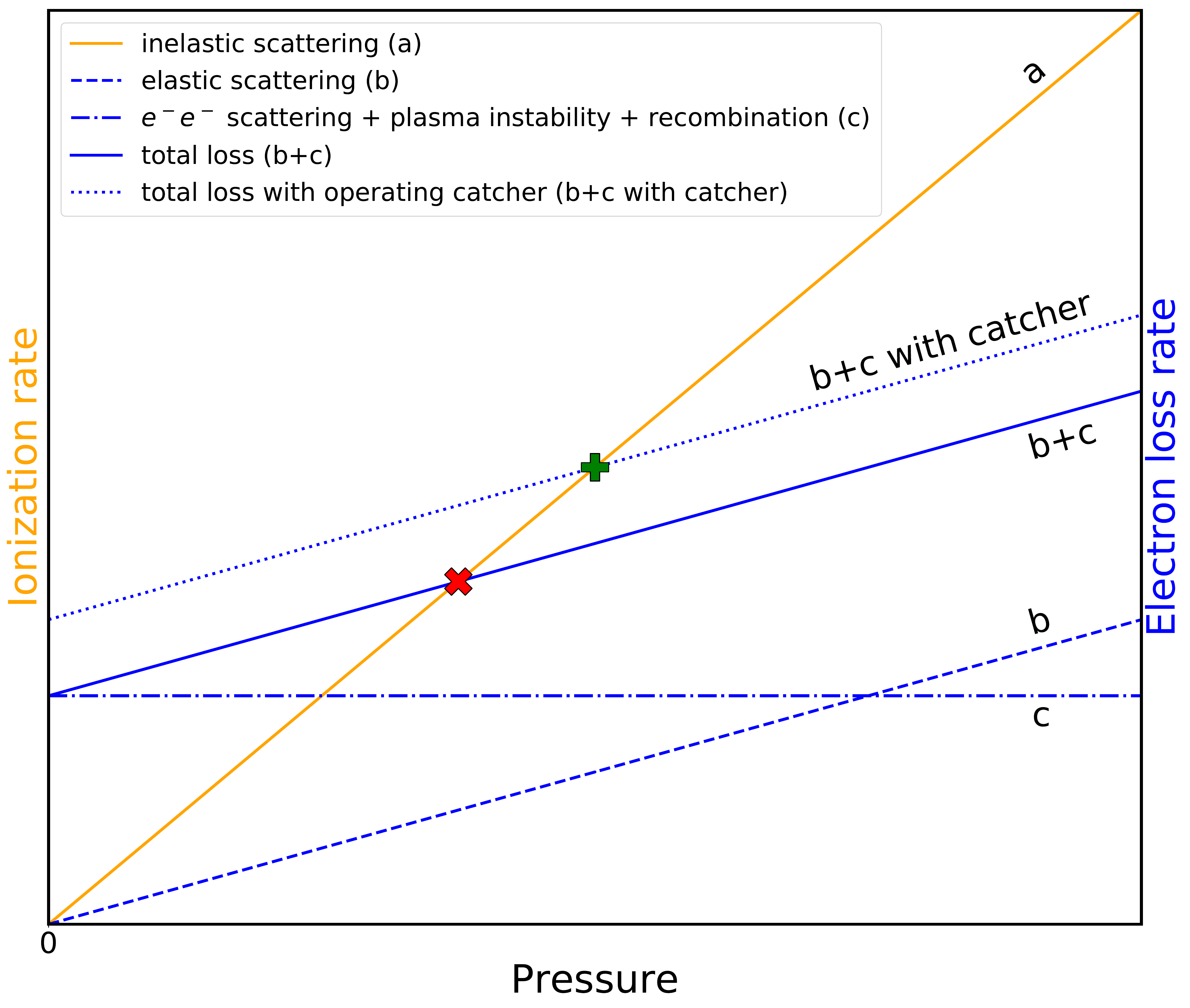}
\caption{A schematic qualitative illustration of the ionization rate due to electron inelastic scattering (solid orange line) and the electron-loss rate (solid blue line).  The latter rate has contributions from elastic scattering (dashed blue line) and Coulomb interaction together with recombination (dash-dotted blue line). 
In the pressure region to the left from the crossing point of these two processes (red cross) the electron-loss rate dominates, preventing the Penning trap from igniting. A continuous operation or a periodic movement of an electron catcher would increase the total electron-loss rate, thereby shifting the corresponding line upwards (dotted blue line) and shifting the crossing point to higher pressures (green plus).  
}
\label{loss_vs_ionization}
\end{figure}

Apart from accumulation and multiplication of electrons, there are possible antagonistic processes which lead to losing electrons from the inter-spectrometer Penning trap. In addition to the elastic scattering on residual gas particles mentioned above, stored electrons will have Coulomb interactions with other electrons and with the plasma; these interactions can promote much more angular changes and thus higher  rates of cyclotron emission. Another possible loss mechanism is due to time-dependent non-axisymmetric electric fields of plasma instability, which can result in electrons leaving the trap in the radial direction. Also, since each of the spectrometers is an electrostatic and magnetic bottle trap for low-energy positive ions, there is a small overlap of the clouds of electrons and ions (stored inside the inter-spectrometer region and the spectrometers, respectively), which can give an additional weak electron-ion recombination contribution. While the rate of elastic scattering on residual gas is directly proportional to its pressure, $\dot{N}_{e}^\mathrm{elastic} = n v \sigma_{e}^\mathrm{elastic}$, the electron-electron scattering, plasma instability, and recombination are pressure-independent. Additionally, for electrons with energies above \SI{90}{eV}, the inelastic cross section is higher than the elastic cross section~\cite{beck}. Therefore, as is schematically illustrated in fig. \ref{loss_vs_ionization}, the electron-loss rate will dominate the ionization rate at low pressures, providing a pressure region where the Penning trap can exist without developing discharges. A continuously operated or periodically actuated electron catcher would additionally remove stored electrons, enlarging the pressure range in which the system can be operated safely.

In view of these studies, the interception of trapped electrons with a movable electron catcher was chosen as the most suitable and reliable method. Therefore, newly developed electron catchers were implemented inside the magnet bore between the two spectrometers. In this paper we report on the successful commissioning of the electron catchers, when their effectiveness for eliminating Penning trap activity was demonstrated.

\section{Working principle and technical realization of the electron catchers}
\label{WorkingPrinciple}

\begin{figure}[ht]
\center
\begin{minipage}[h]{\columnwidth}
\center
\begin{minipage}[h]{\columnwidth}
\center{\includegraphics[width=\columnwidth]{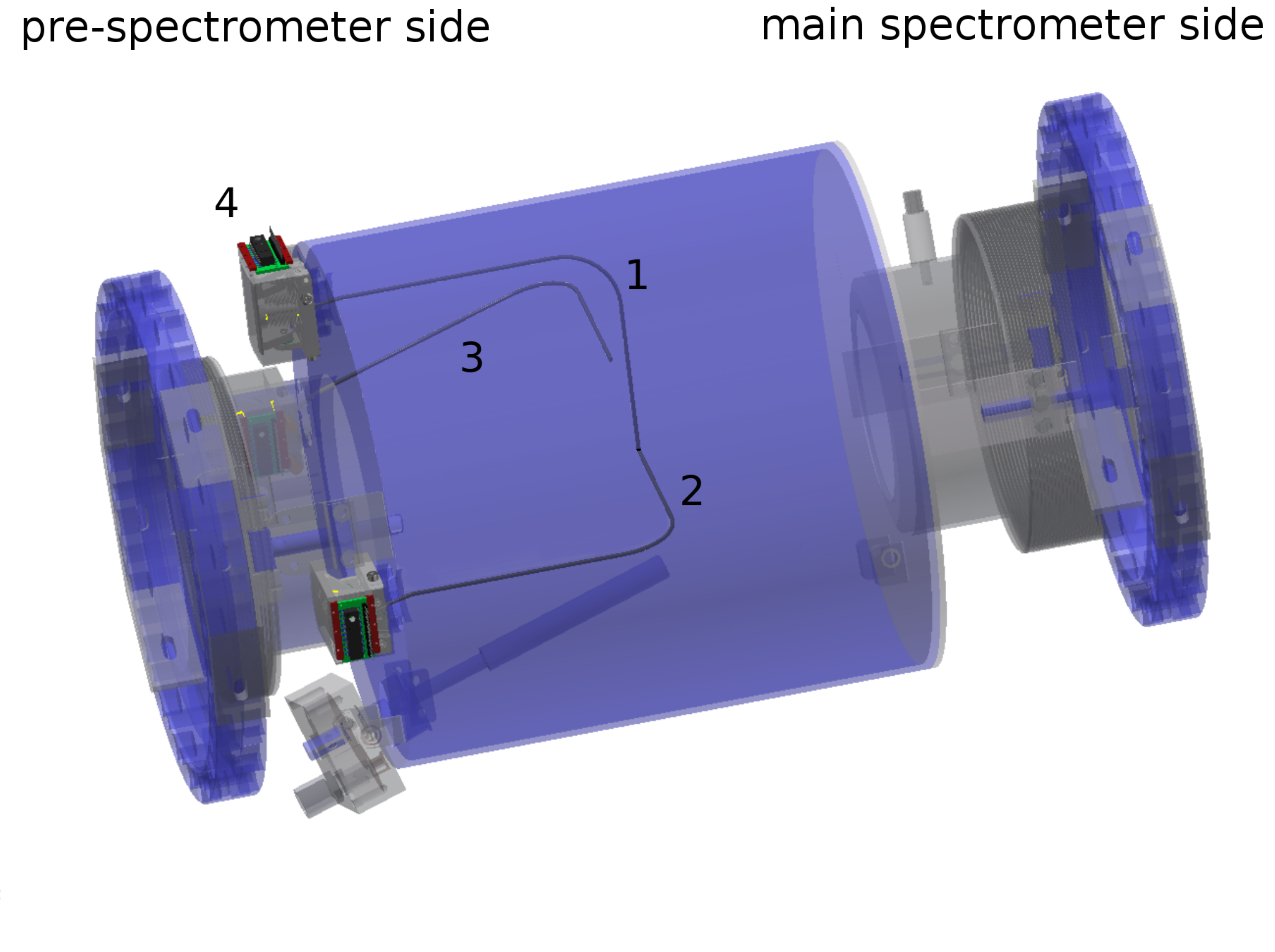} \\ a)}
\end{minipage}
\vfill
\begin{minipage}[h]{\columnwidth}
\center{\includegraphics[width=\columnwidth]{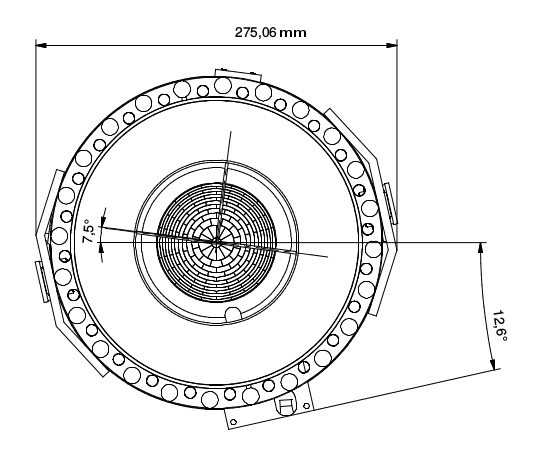} \\ b)}
\end{minipage}
\caption{a) The beamline region inside the superconducting solenoid between the pre- and main spectrometers. (1), (2), (3) - the electron catchers; (4) - a catcher's controller; b) End view of the electron catchers in their operational positions, relative to the FPD looking down the beamline toward the main spectrometer.}
\label{PW_drawing}
\end{minipage}
\end{figure}

The three electron catchers are bent rods (\SI{2}{mm} in diameter) made of nickel-chromium-iron alloy, Inconel, and installed at three different positions inside the valve connecting the pre- and main spectrometers, as shown in fig. \ref{PW_drawing}. Each of the catchers is connected at one end to an individual bellows attached to the valve housing. This movable connection allows the free end of the catcher to be moved from its parking position (outside the flux tube) to its operating position (within the flux tube). When it is moved in, it traverses the flux tube in a radial direction from its edge up to the center. The catchers are tilted in such a way (\SI{7.5}{^\circ} with respect to the horizontal axis; see fig. \ref{PW_drawing}b) that the number of detector pixels affected when the catcher is inside the flux tube is minimized.

The principle of electron removal with the installed electron catchers is based on the specifics of electron motion inside the Penning trap. In a classical Penning trap \cite{blaum}, electrons are constrained radially by a homogeneous axial magnetic field and axially by a quadrupole electric field. In a homogeneous magnetic field, an electron would move in a circular motion, called the cyclotron motion, around the magnetic field lines with a corresponding cyclotron frequency:

\begin{equation}
	\omega_c = \frac{q}{m} B.
	\label{cyclotron}
\end{equation}

Due to the additional electric field in the Penning trap and due to inhomogeneities of the magnetic fields, stored electrons undergo a more complex motion, which can be represented as a superposition of three different components: 
\begin{enumerate}
\item the axial oscillation along the trap axis, independent of the magnetic field, with the frequency

\begin{equation}
	\omega_z = \sqrt{\frac{q}{m}\frac{U}{d^2}},
	\label{axial}
\end{equation}

where $U$ is called the ``trap depth", as particles with an energy lower than $qU$ will not be able to escape the trap. The parameter $d$ describes the trap dimensions, and in the case of a classical Penning trap with hyperbolic electrodes it is

\begin{equation}
	d^2 = \frac{1}{2} \left( z_0^2 + \frac{r_0^2}{2} \right),
	\label{d}
\end{equation}

where $2z_0$ is the distance between electrodes and $r_0$ is the trap radius.
\item the ``modified" cyclotron motion, which has a frequency slightly different from eqn.~\eqref{cyclotron} but which strongly depends on the trap depth and dimensions:

\begin{equation}
	\omega_+ = \frac{\omega_c}{2} + \sqrt{\frac{\omega_c^2}{4} - \frac{\omega_z^2}{2}} \approx \omega_c - \frac{U}{2d^2B}.
\end{equation}

\item the magnetron drift around the trap axis (formula given for a homogeneous magnetic field):

\begin{equation}
	\omega_- = \frac{\omega_c}{2} - \sqrt{\frac{\omega_c^2}{4} - \frac{\omega_z^2}{2}} \approx \frac{U}{2d^2B}.
\end{equation}

\end{enumerate}

\begin{figure}[ht]
\center
\includegraphics[width=\columnwidth]{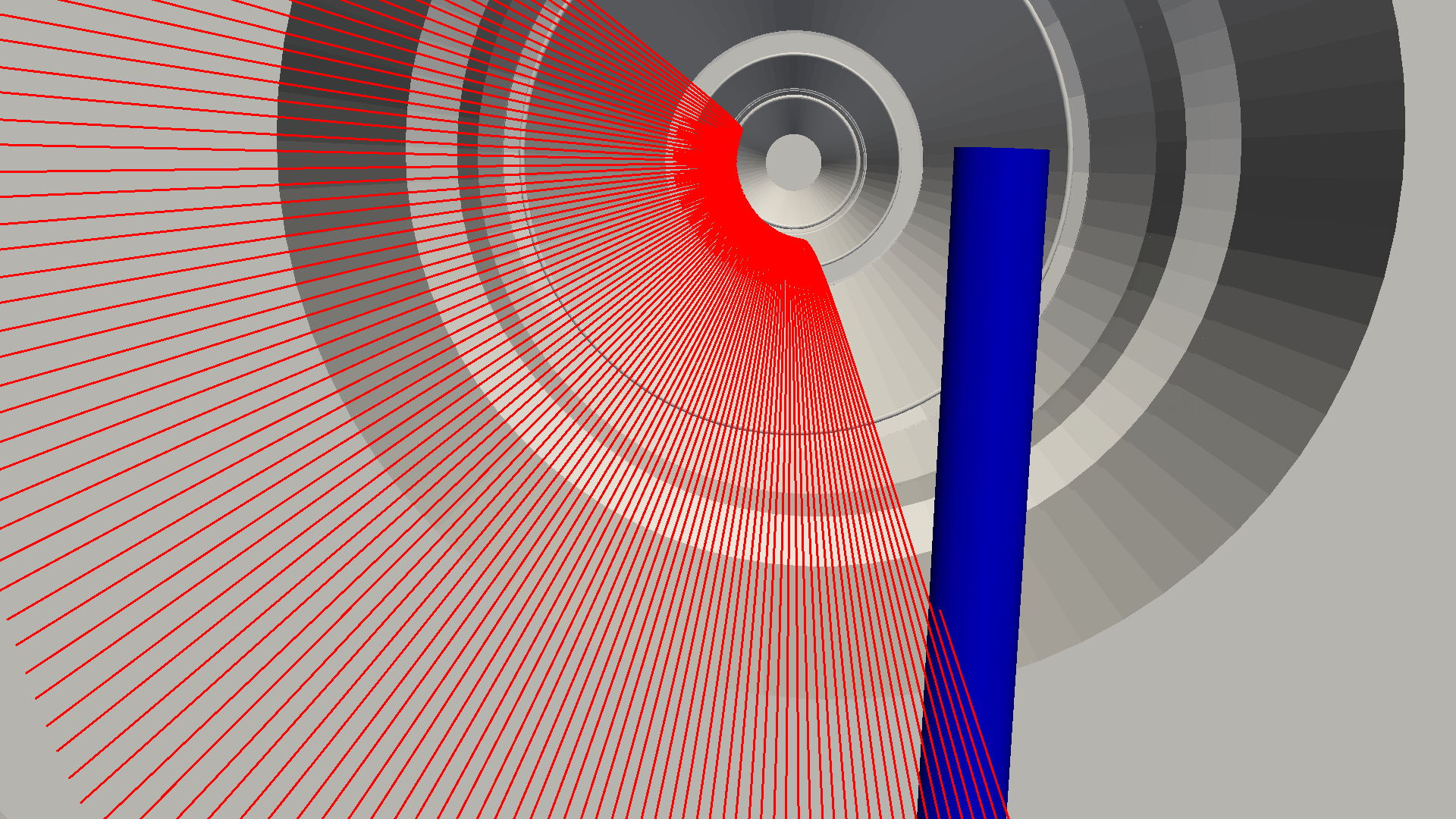}
\caption{Simulated track of a trapped electron (red) up to the point where it hits the electron catcher (blue) (looking down the beamline). Simulation is done with the Kassiopeia software \cite{Kassiopeia}.}
\label{simulation_magnetron}
\end{figure}

In the case of the inter-spectrometer Penning trap, which has a length of  $2z_0\approx\SI{1.5}{m}$ (see fig. \ref{PT_fields}) and a trap radius of $r_0=\SI{3,6}{cm}$ at the center of the solenoid (due to the confinement of the transported magnetic flux\footnote{which is constant over the entire KATRIN beamline} of $\phi= \SI{133.7}{Tcm^2}$ \cite{KATRIN_design} by the magnetic field of \SI{3.2}{T}), the corresponding dimensional parameter is $d \approx\SI{1.1}{m}$ (according to eqn.~\eqref{d}).
Because the trap depth is about $U=\SI{-18}{kV}$, the frequencies of the stored electrons are $\omega_z \approx\SI{5e7}{s^{-1}}$, $\omega_+ \approx\SI{8e11}{s^{-1}}$ and $\omega_- \approx\SI{2e3}{s^{-1}}$. Due to magnetron motion, which is slow with respect to the axial and cyclotron components but still fast compared to the subsecond-scale mechanical movement of the catcher, an electron stored somewhere inside the trap will eventually be intercepted by the electron catcher. Faster axial motion will lead to ``gaps" along the circumference of the electron's magnetron motion, but the size of the gaps is much smaller than the diameter of the electron catchers, so the trapped electron will not be able to avoid the catcher (see fig.~\ref{simulation_magnetron}). Each catcher is designed such that when it is inserted into the flux tube its free end reaches the center of the flux tube. This guarantees that all stored electrons will in the end be removed by hitting the inserted catcher. As has been discussed above, for a pressure of \SI{e-11}{mbar} the time interval between ionizations in the Penning trap is of the order of tens of seconds and the electron cooling time is of the order of \SI{1}{h}. Because these timescales are much longer than the sub-ms scale of the magnetron motion, the catcher is able to stop the electron multiplication process and effectively suppress the corresponding background \footnote{Note that these timescale considerations are only valid under the condition that the trapped electron does not interact with other accumulated electrons, which does not hold when a Penning discharge takes place.}.

\begin{figure}[ht]
\center
\includegraphics[width=\columnwidth]{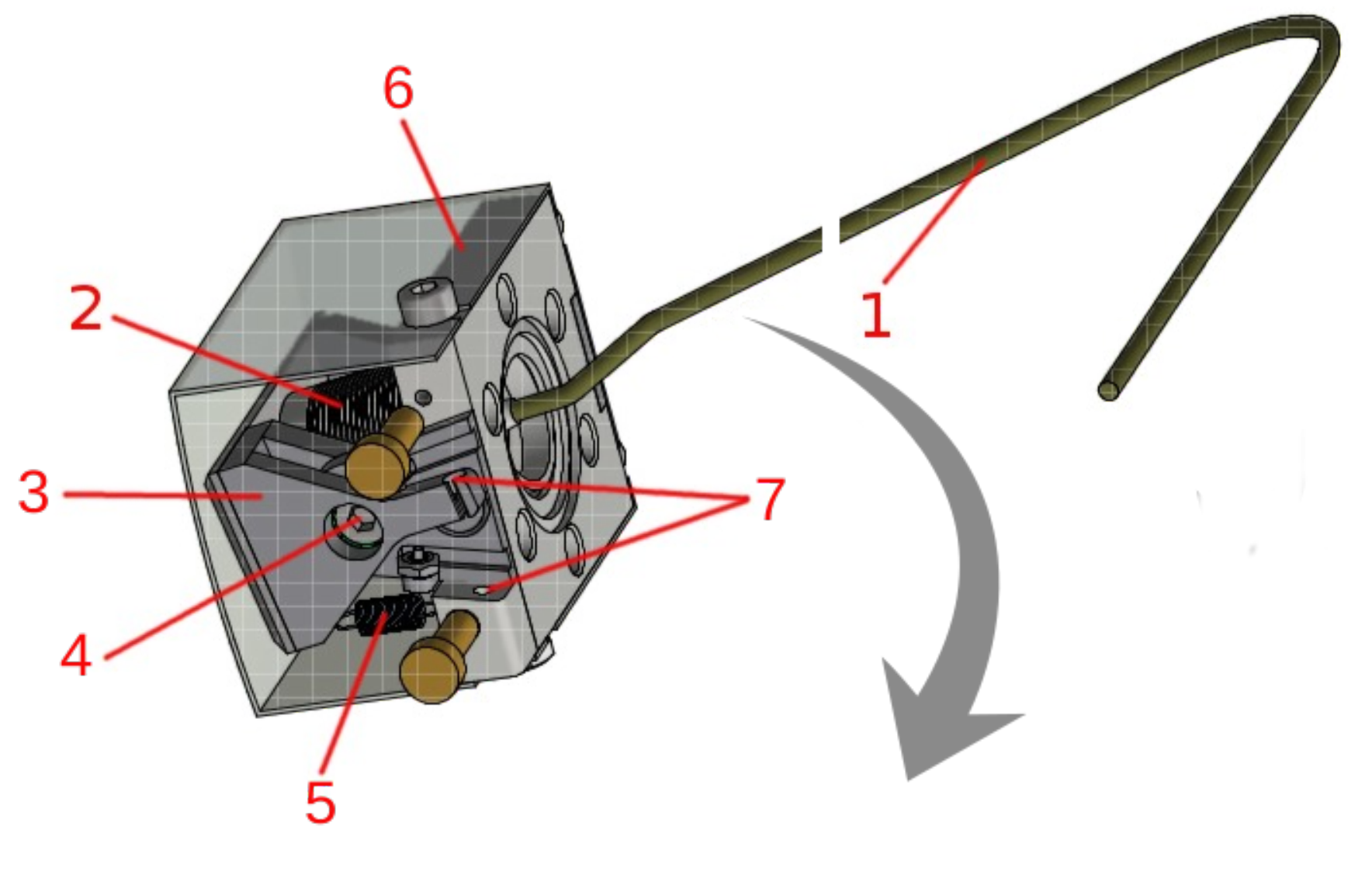}
\caption{Technical drawing of an electron catcher controller: (1) - electron catcher, (2) - bellows, (3) - hinge, (4) - rotation axis, (5) - spring, (6) - photosensor housing, (7) - pneumatic muscle cable slots. The grey arrow shows the direction of the catcher when moving inside the flux tube.} 
\label{catcher_drawing}
\end{figure}

The mechanical design of the electron catchers is shown on fig. \ref{catcher_drawing}. The catcher is attached to a bellows fixed inside a hinge to allow for tilting around one axis. This joint is contained within a housing, and a spring attached to one of the walls holds the hinge in the parking position. On the other side, the hinge is connected via a Bowden cable to a pneumatic muscle which, when contracted, pulls the hinge and moves the catcher into the flux tube. The pressurized air actuating the muscle is supplied through a pneumatic valve from a compressed air supply. The contraction speed of the pneumatic muscle, which determines the electron catcher sweeping speed (one-way movement on a sub-second scale), can be regulated by manually adjusting a flow control valve. The photoelectric sensor of each electron catcher contains an infrared LED and a photodiode. When the catcher is moved into the flux tube, the LED light is reflected from the hinge and strikes the photodiode. All this instrumentation was developed and tested to be fully compatible within a multi-Tesla magnetic field.

\begin{figure}[ht]
\center
\includegraphics[width=\columnwidth]{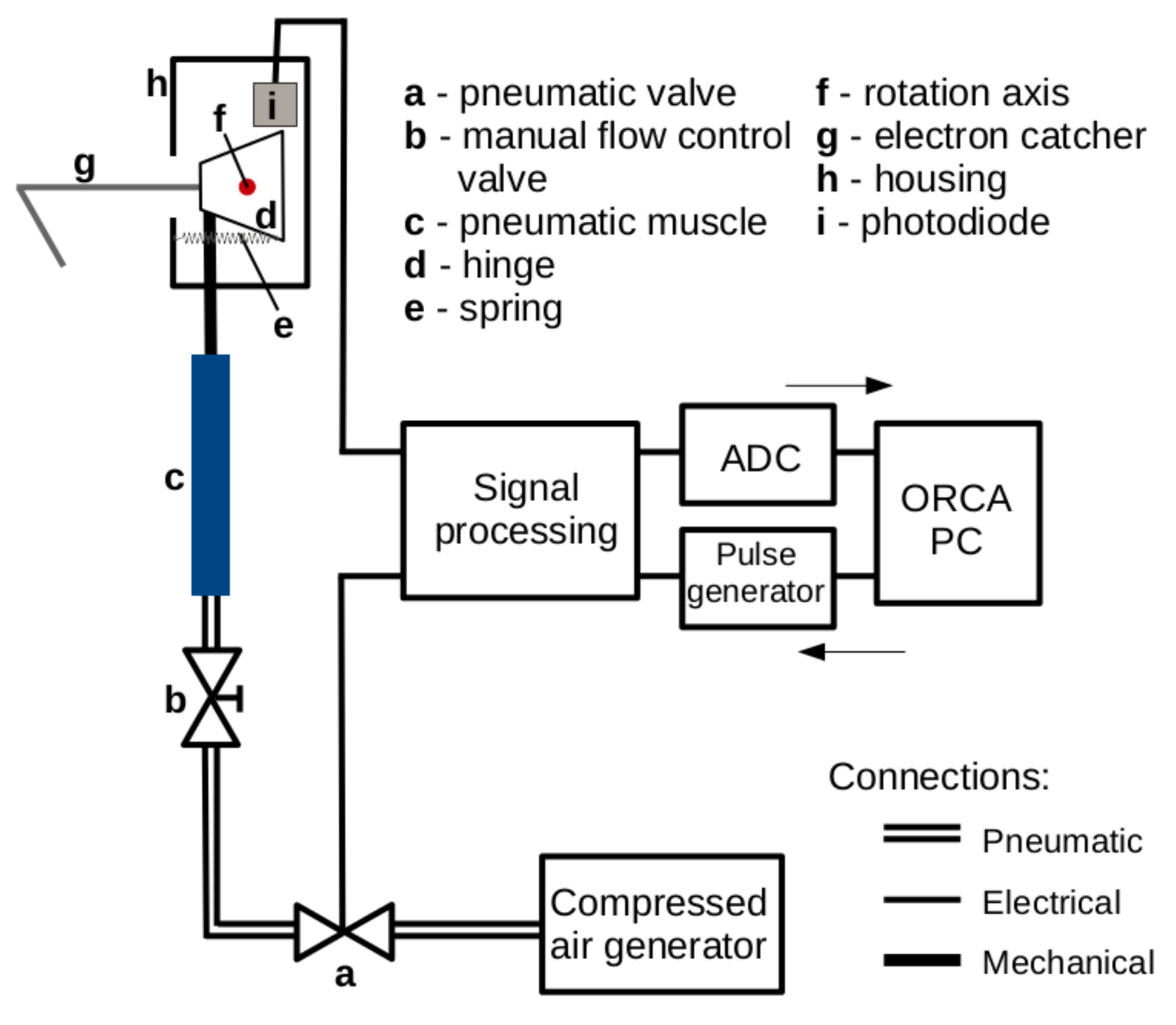}
\caption{Schematic diagram of the electron catcher operation. The arrows show the direction of the data flow. To remove the catcher from the flux tube: 1) the pulse generator is first activated by ORCA and creates a TTL signal, 2) the signal proceeds to the signal processing stage and is transformed to a 24-V signal which closes the pneumatic valve (a) to the compressed air supply, 3) the pneumatic muscle (c), which is connected to the hinge (d) held by the spring (e), is relaxed. When the signal drops low, the valve reopens, causing the contraction of the muscle which then pulls the hinge and rotates it around the axis (f), thereby moving the electron catcher (g) back to the flux tube. The speed of the muscle contraction and consequently of the catcher's movement can be regulated by the manual flow control valve (b). In this configuration, infrared light from an LED installed inside the photoelectric housing (h) is reflected from the hinge onto the photodiode (i). The photodiode signal enters the signal processing stage and is converted to a TTL signal. The signal increase (or decrease for the case in which the catcher is removed from the flux tube) is read out by the DAQ system.}
\label{moving_setup}
\end{figure}

A schematic illustration of the combined setup used to operate a single electron catcher is shown in fig. \ref{moving_setup}. 
The system works in a triggering mode: a TTL signal is created by a pulse generator (Agilent 33220A) and is converted at the signal processing stage to a 24-V signal. As long as the signal stays at this level the pneumatic valve to the muscle is closed, the muscle is relaxed, and the catcher remains outside the flux tube. When the signal drops low the valve is opened, the muscle contracts, and the catcher moves into the flux tube. In this case the photodiode creates a current that is sent to the signal processing stage where it is converted to a TTL signal. The rising edge of the signal (or its trailing edge in the opposite case when the catcher is moving out of the flux tube) is detected, read out, and synchronized with the ORCA data stream. The pulse generator controlled by ORCA allows one to adjust the movement patterns (``in" and ``out" times) and to operate the catcher in different modes: a) automatic mode, in which the catcher is inserted into the flux tube at regular, adjustable intervals, b) safeguard mode, in which the catcher is triggered when the detected electron rate at the FPD surpasses a user-defined threshold, and c) manual operation mode.

\begin{figure}[ht]
\center
\begin{minipage}[h]{\columnwidth}
\center
\begin{minipage}[h]{\columnwidth}
\center{\includegraphics[width=\columnwidth]{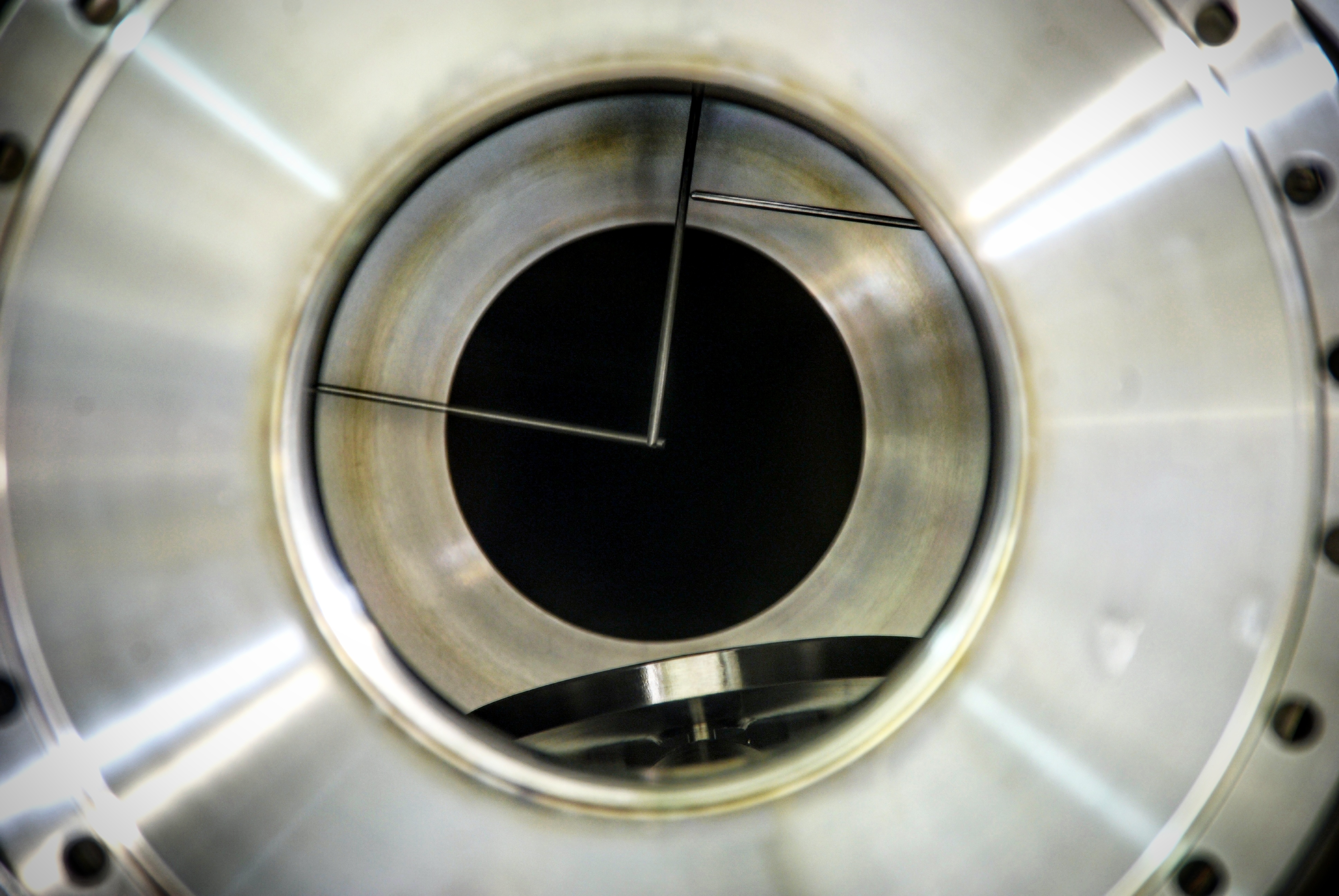} \\ (a)}
\end{minipage}
\vfill
\begin{minipage}[h]{\columnwidth}
\center{\includegraphics[width=\columnwidth]{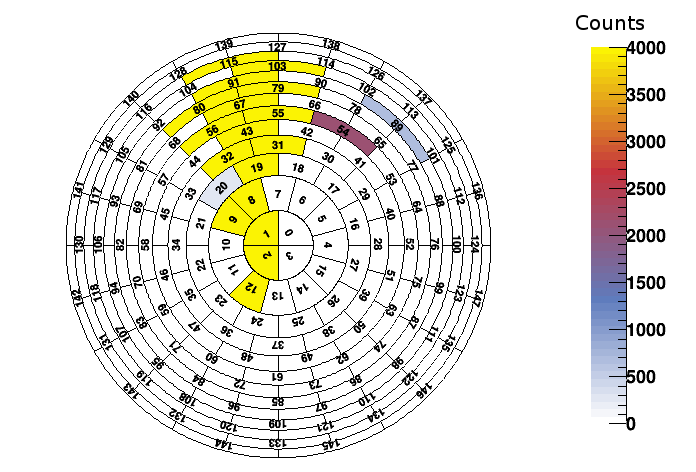} \\ (b)}
\end{minipage}
\caption{(a): View of the catchers installed in the valve between the spectrometers (looking down the beamline). The left and central catchers are inserted into the flux tube, while the right catcher is retracted. (b): FPD pixels affected by the shadow from the central electron catcher (the position of the shadow is mirrored to the image of the catcher on (a) because of the opposite direction of view). The total electron count from a photoelectron source was measured with the catcher inserted into the flux tube, and this count was then subtracted from the count obtained with all three catchers retracted for the same amount of time (10 min); therefore, higher counts are observed for the shadowed pixels.}
\label{PW_against_FPD}
\end{minipage}
\end{figure}

A continuously inserted pin would shadow some of the FPD pixels (see fig. \ref{PW_against_FPD}). Therefore, the ability to quickly move the electron catchers into and out of the flux tube prevents unnecessary losses in statistics. The bellows of the catchers are specified to 1.5 million movements, which makes a total of 4.5 million movements for the three catchers, which in turn allows to move the catchers about every 20 s during the 3 years of measurement time of KATRIN.

In the following section, we report on our investigations of the inter-spectrometer Penning trap behavior and the efficiency with which the electron catchers empty the trap.

\section{Measurements and results}

\subsection{\textit{Background measurements before bake-out}}

\begin{figure}[ht]
\center
\includegraphics[width=\columnwidth]{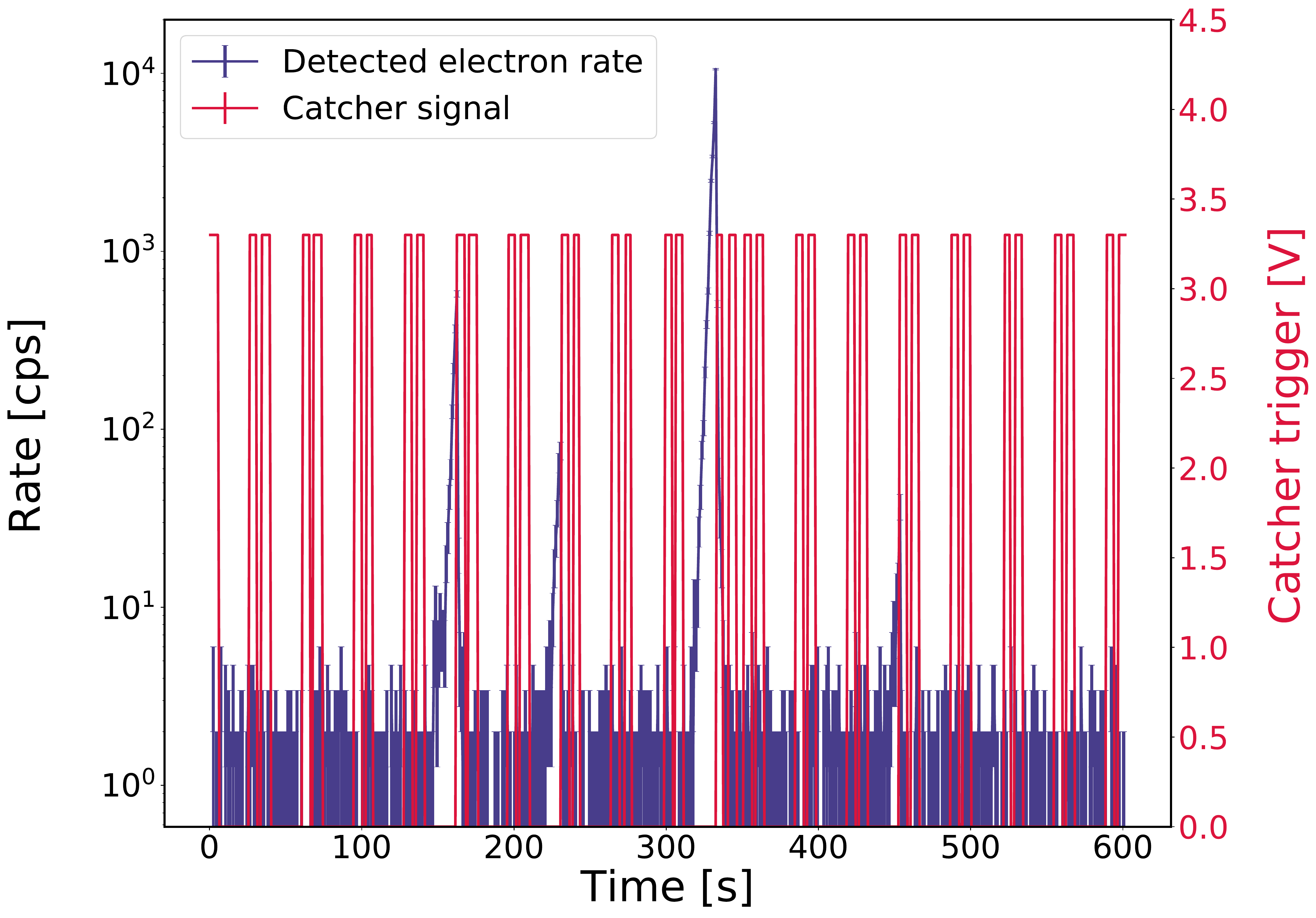}
\caption{Focal Plane Detector background electron rate for setting A, table \ref{tab:before_after_bake_out} and 
the electron catcher operated in a mode in which it moved in a repeated pattern of \SI{4}{s} inside the flux tube, \SI{4}{s} outside the flux tube, \SI{4}{s} inside the flux tube, and \SI{30}{s} outside the flux tube; however, when the safeguard script was triggered by an FPD rate higher than \SI{10}{kcps}, the catcher was inserted an additional two times (each time for \SI{4}{s}, with \SI{4}{s} in between).}
\label{FL+_moving-wiper}
\end{figure}

\begin{figure}[ht]
\center
\includegraphics[width=\columnwidth]{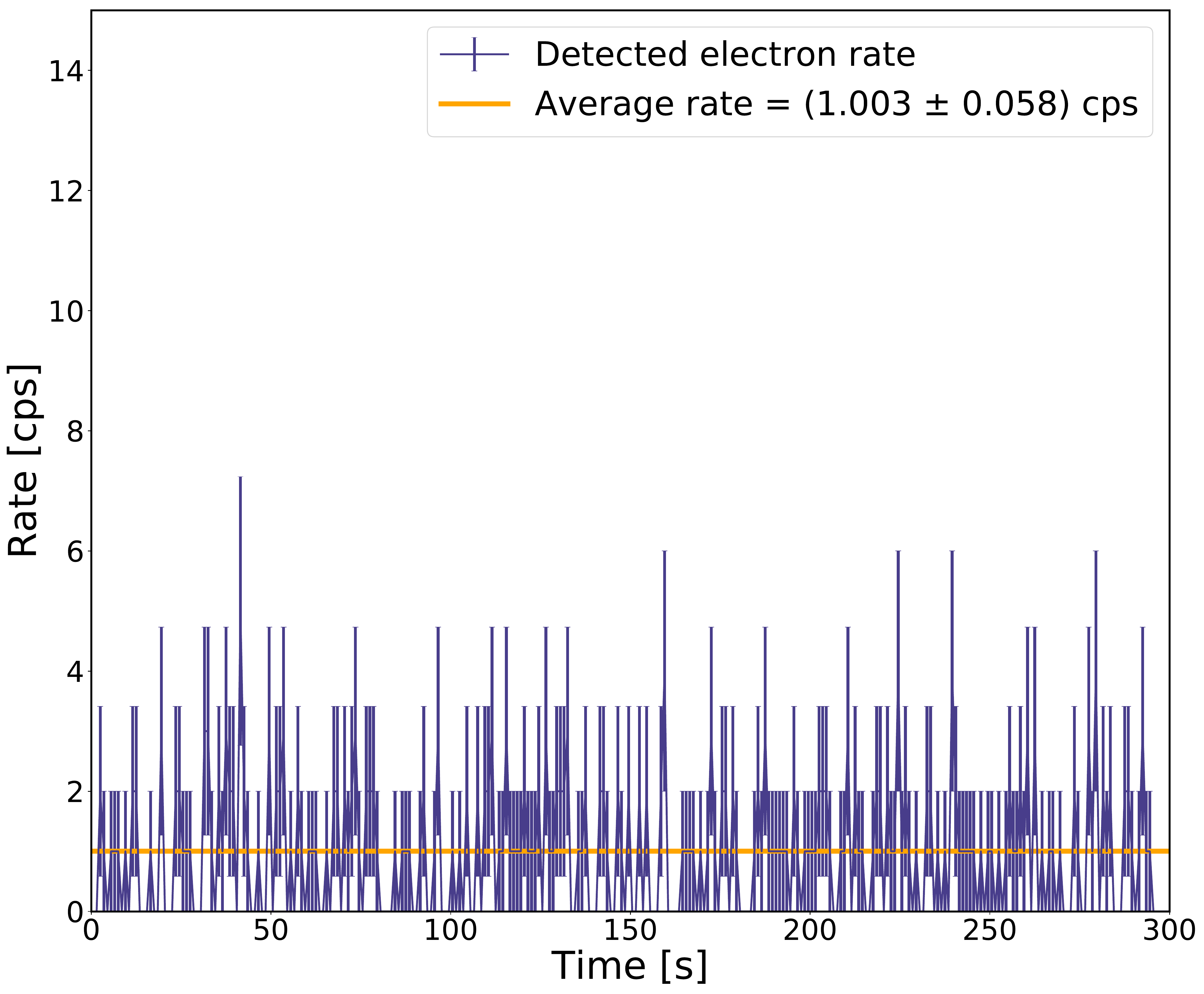}
\caption{Focal Plane Detector background electron rate for setting A, table \ref{tab:before_after_bake_out} and a stationary electron catcher inside the flux tube. No rate bursts were observed with this configuration.}
\label{FL+_static-wiper_32148}
\end{figure}

During the first test of the system in 2016, the pressure inside the spectrometers was on the order of \SI{e-9}{mbar}. With the electron catchers retracted from the flux tube, the main spectrometer was operated at $U_{\mathrm{MS}}=$\SI{-18.6}{kV}, and the beamline magnetic field was set to 20$\%$ of its maximum value (setting A, table \ref{tab:before_after_bake_out}).  At a pre-spectrometer voltage of $U_{\mathrm{PS}}=$\SI{-2.6}{kV}, Penning discharges appeared within \SI{5}{min} from the start of the background measurement. With the magnetic field set to 80$\%$ of its maximum value (setting B, table \ref{tab:before_after_bake_out}), this voltage threshold value was even lower, at $U_{\mathrm{PS}}=$\SI{-2.0}{kV}, consistent with the fact that higher magnetic fields create stronger trapping conditions. Frequent and intense rate bursts were observed unless stopped by an electron catcher. Figure \ref{FL+_moving-wiper} shows a test in which one of the electron catchers was moved into the flux tube in a regular pattern with 30-second breaks, during which background spikes developed a couple of times. The electron catchers were tested in different operational modes. A periodically actuated electron catcher demonstrated the ability to quench erupting discharges, but it couldn't prevent their full development. A continuously operated catcher made it possible to reach \SI{-18.4}{kV} at the pre-spectrometer without observing background rate bursts (see fig. \ref{FL+_static-wiper_32148}). However, this configuration is disadvantageous for data-taking during neutrino mass measurements, since a non-negligible part of the measurement statistics would be lost due to pixels shadowed by the catcher.

\begin{table*}[t]
\begin{threeparttable}
%\resizebox{1.\textwidth}{!}{
\begin{tabular*}{\textwidth}{c @{\extracolsep{\fill}} c c c c c c c c c c}

\hline

\textbf{Setting} & \textbf{A} & \textbf{B} & \textbf{C}& \textbf{D}& \textbf{E}& \textbf{F}& \textbf{G}& \textbf{H}\\

\hline

\textbf{Pressure, [mbar]} & $\sim 10^{-9}$ &	$\sim 10^{-9}$ &	  variable &	variable &	$ \sim 10^{-11}$ &  $ \sim 10^{-11}$ & $ \sim 10^{-11}$ & $ \sim 10^{-11}$		\\
\textbf{Solenoids' magnetic field} &  20$\%$ & 80$\%$	& 70$\%$	& 70$\%$	& 70$\%$\tnote{\dag}	& 	70$\%$ & 70$\%$ 	& 70$\%$		\\
\textbf{LFCS magnetic field} &  20$\%$ & 80$\%$	& 42$\%$	& 42$\%$	& 70$\%$	& 42$\%$	& 42$\%$	&	70$\%$	\\
\textbf{$\bm{U_{\mathbf{MS}}}$, [kV]} &		$-18.6$ &$-18.6$ &$-18.6$ & $-15.1$	&	$-18.6$ &	$-18.6$ &	$-18.6$ & $-18.6$		\\
\textbf{$\bm{U_{\mathbf{PS}}}$, [kV]} &		variable 	&variable 	&variable 	& $-18.4$	& $-18.3$	& $-18.3$	& $-18.3$	& $-18.33$		\\
\textbf{Ion gauge} &	 On	&On	&On	&	Off  & Off	& On	& Off	& Off		\\
\hline
\end{tabular*}
%}
\caption{Settings for measurements before (A, B) and after (C-H) the bake-out of the spectrometers. The magnetic fields are given in percentage of the maximal field settings for the the spectrometers' solenoids (e.g. \SI{4.5}{T} for the common magnet) and the LFCS (\SI{0.9}{mT}).}
\label{tab:before_after_bake_out}
\begin{tablenotes}
\item[\dag] Except for the pre-spectrometer magnets operated at \SI{4.1}{T} to mimic the influence of magnetic stray fields from the transport section in the main spectrometer.
\end{tablenotes}
\end{threeparttable}
\end{table*}


\subsection{\textit{Background measurements at different pressures after bake-out}}

\begin{figure}[ht]
\center\includegraphics[width=\columnwidth]{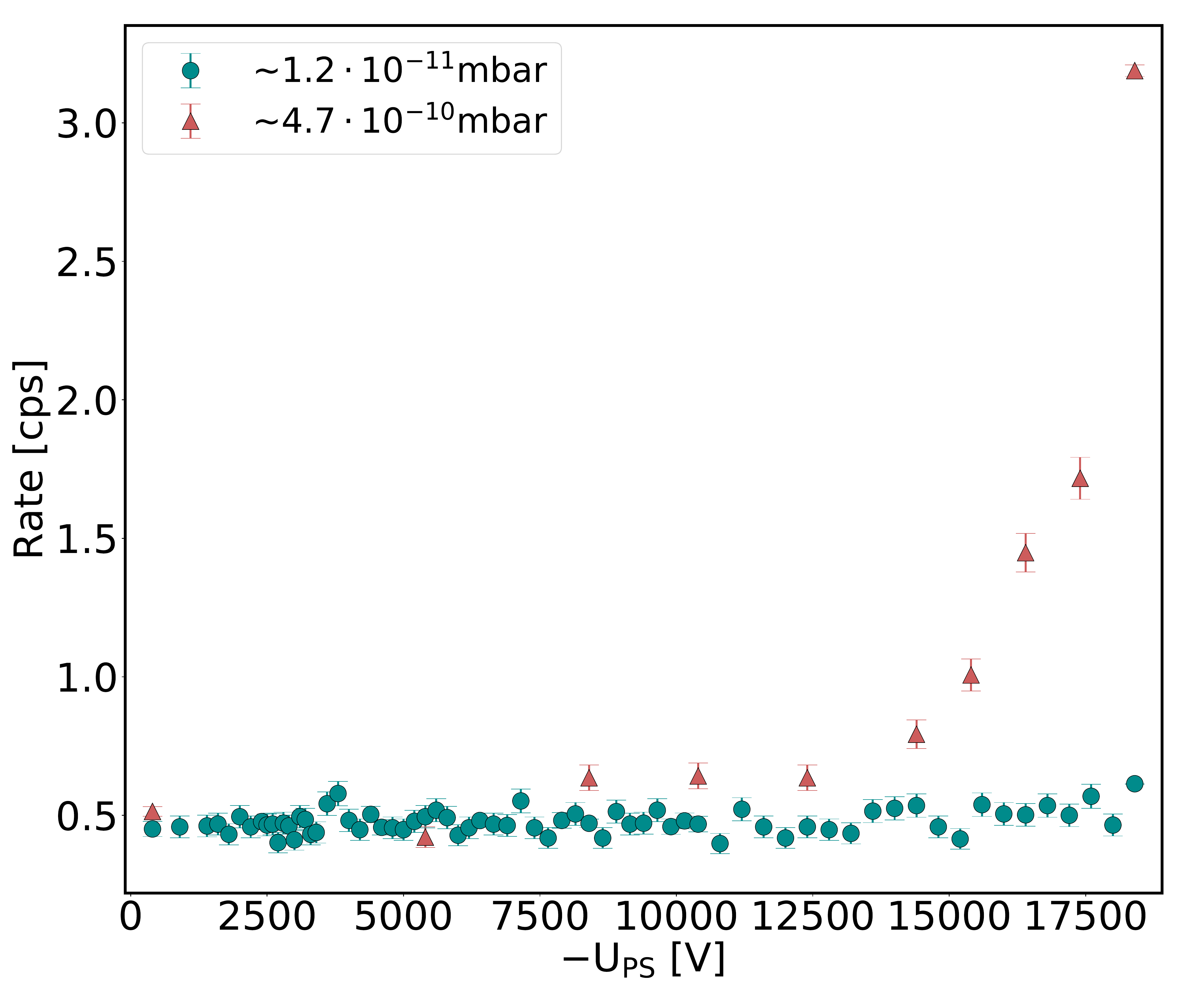}
\caption{Background measurements at two different pressures and variable $U_{\mathrm{PS}}$ (setting C, table \ref{tab:before_after_bake_out}). A strong increase of the background rate induced by the Penning trap manifests itself during the higher pressure measurements (red dots) for PS voltages more negative than \SI{-14.4}{kV}, while for the lower pressure measurements (blue dots) the rate stays essentially unchanged for the whole range of $U_{\mathrm{PS}}$. For the measurement at increased pressure, the indicated pressure value is corrected for argon.}
\label{pressure_dependence}
\end{figure}

Before the subsequent measurement in 2017, the system was baked out \cite{vacuum}, and tests were conducted at different pressures. The pressure in the system was changed by injecting small amounts of argon gas through one of the main spectrometer pump ports. Two sets of measurements, one at a residual gas pressure of $\sim$\SI{1.2e-11}{mbar} and the other at $\sim$\SI{4.7e-10}{mbar} (setting C, table \ref{tab:before_after_bake_out}), were performed to determine the background rate. As can be seen in fig. \ref{pressure_dependence}, the system was very sensitive to the pressure: while the rate was low and stayed constant at the pressure of \SI{1.2e-11}{mbar}, a dramatic increase was observed when the pressure was increased fortyfold and the pre-spectrometer was ramped to more negative potentials below \SI{-14.4}{kV}.

\begin{figure}[ht]
\center
\includegraphics[width=\columnwidth]{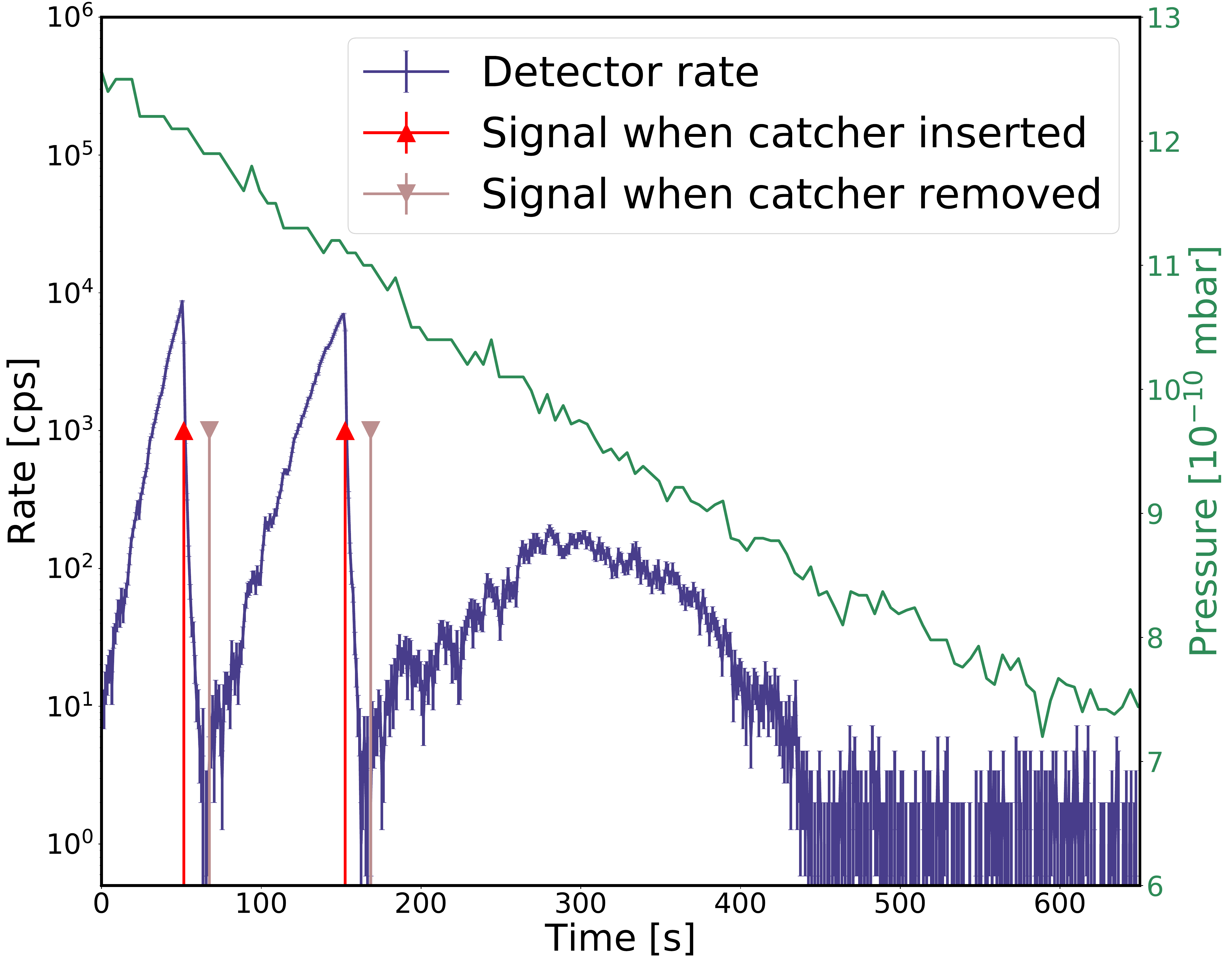}
\caption{Background measurement with decreasing pressure (by pumping out previously injected argon gas) when the electron catcher was run in safeguard mode (setting D, table \ref{tab:before_after_bake_out}). The electron catcher successfully quenches discharges until the discharges disappear below $\sim$\SI{8e-10}{mbar}. Pressure values are corrected for argon.}
\label{SDS-III_moving-wiper_34342}
\end{figure}

During the same measurement, the catcher was tested in its safeguard mode at elevated pressures. The ORCA software was set to trigger the electron catcher for \SI{20}{s} when the FPD rate surpassed \SI{10}{kcps}. In fig. \ref{SDS-III_moving-wiper_34342} we show the background rate as the pressure fell from $\sim\SI{1.3e-9}{mbar}$ (when regular discharges were observed and effectively suppressed by the electron catcher) to $\sim\SI{7e-10}{mbar}$ (when the discharges disappeared completely) by pumping out previously injected argon gas. 

\subsection{\textit{Background measurements at nominal pressure and spectrometers' settings}}

\begin{figure}[ht]
\center
\includegraphics[width=\columnwidth]{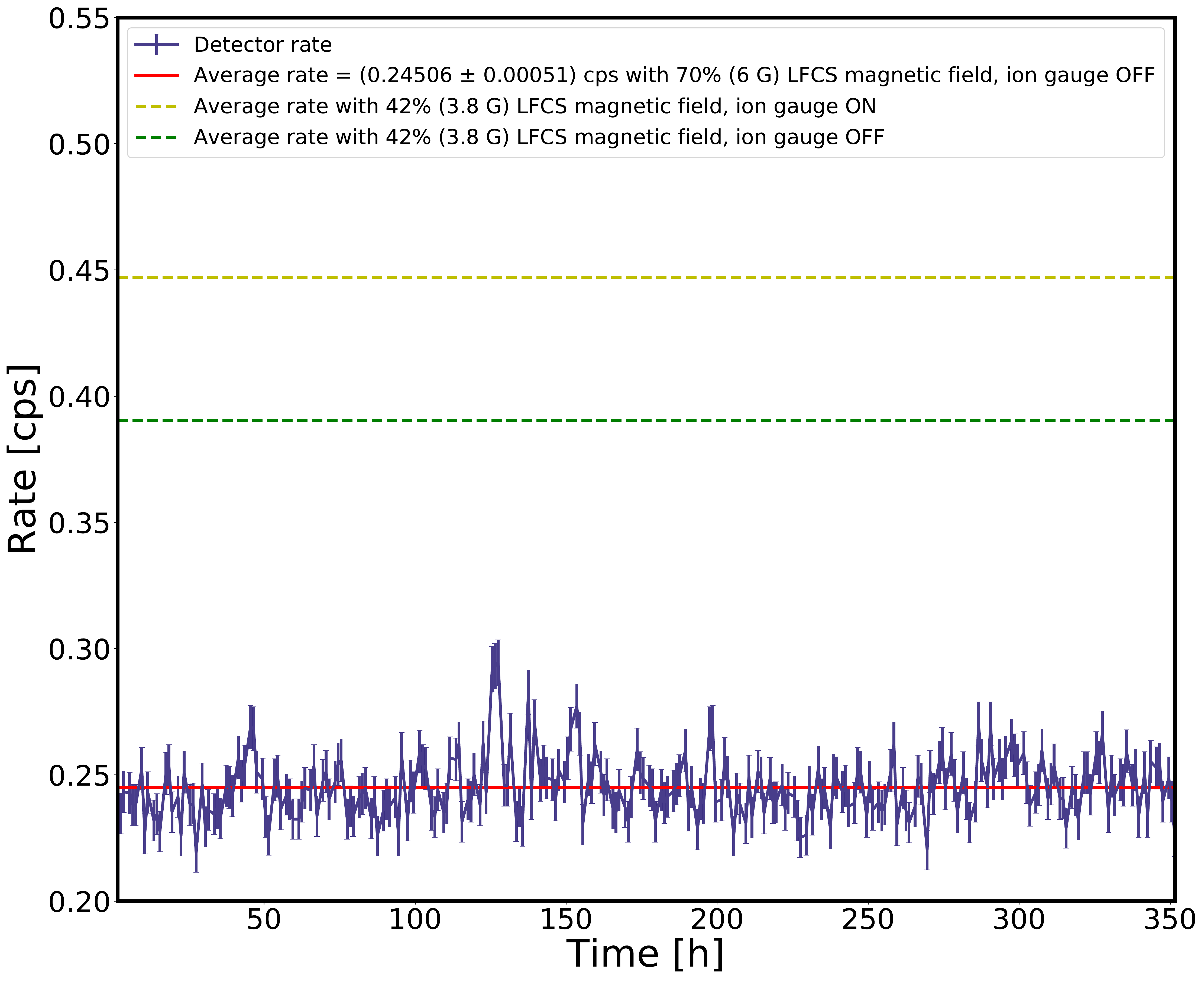}
\caption{Background rate during a long-term measurement at nominal spectrometer conditions (setting E, table \ref{tab:before_after_bake_out}). Shown for comparison are the average rates from measurements taken with a lower LFCS magnetic field in which the ion gauge state was toggled (settings F and G, table \ref{tab:before_after_bake_out}). For all three measurements here, two pixel rings (pixels 4-15 and 136-147, see fig. \ref{PW_against_FPD}b) which were damaged by a Penning discharge during one of the tests before the long-term measurement are excluded.}
\label{Xmass_35023-35110_comparison}
\end{figure}

The bake-out of the spectrometers allowed them to reach their nominal pressure of $\sim\SI{e-11}{mbar}$. The electric and magnetic field settings at the spectrometers planned for use during the neutrino mass measurements were used in a long-term background measurement in late 2017. No discharges or rate spikes were observed during two weeks of data-taking (see fig.\ref{Xmass_35023-35110_comparison}; here, two pixel rings (pixels 4-10 and 136-147, see fig. \ref{PW_against_FPD}b) which were damaged by a Penning discharge during one of the tests before the long-term measurement are excluded). For these measurements, the ion gauge at the pre-spectrometer was switched off since it was previously shown to be a background source. The electron catcher system was operated in safeguard mode during the measurements but was not triggered.

\subsection{\textit{Measurements with active tritium source}}

The observation that no Penning discharge appeared during normal operation of the KATRIN spectrometers at a pressure of $10^{-11}$ mbar does not mean that the Penning trap was completely empty and thus inactive. We only can conclude that in this low-pressure regime the loss processes in the Penning trap were faster than the ionization processes. This prohibits the build-up of an avalanche of stored electrons.
However, if a huge flux of electrons was sent into the pre-spectrometer, some residual effects from the Penning trap would still be expected.

\begin{figure}[ht]
\center
\includegraphics[width=\columnwidth]{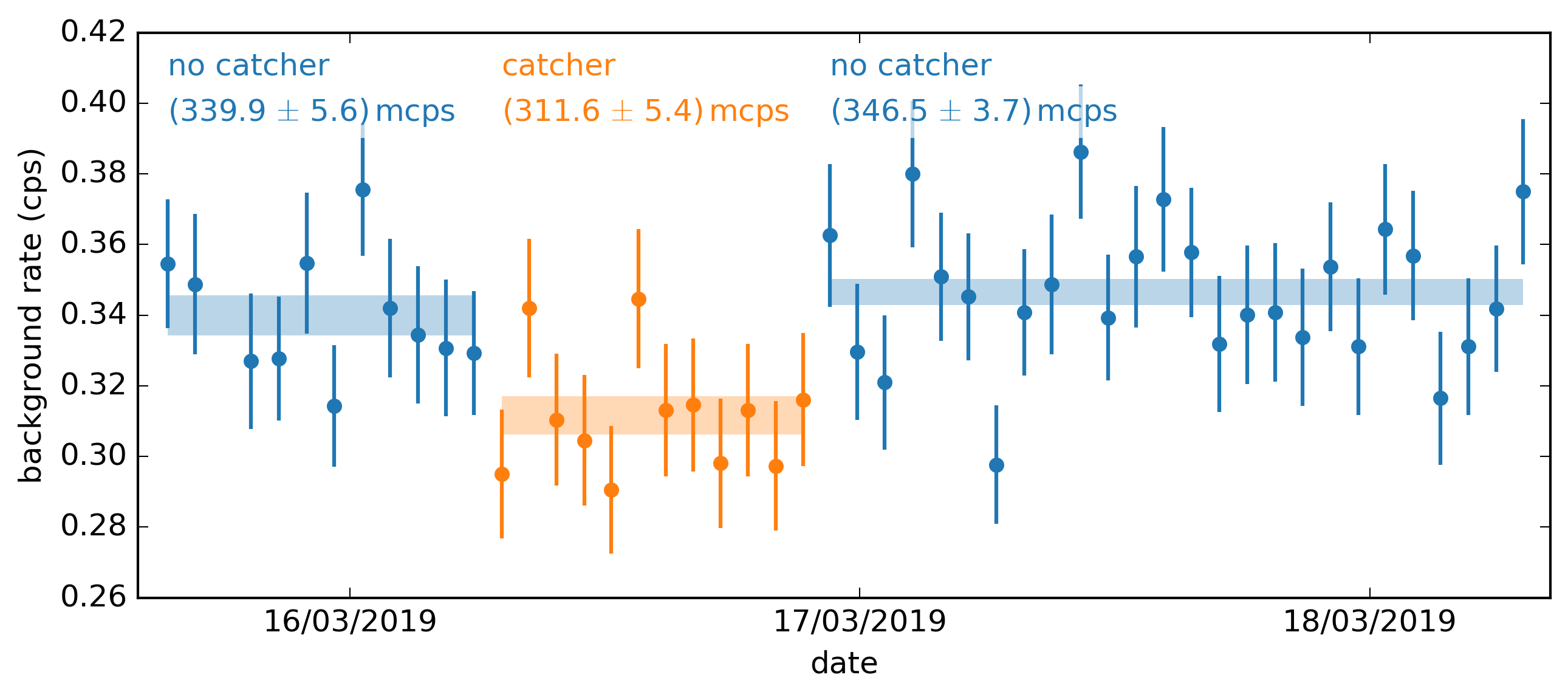}
\caption{Electron rate during measurements with about $2.5 \cdot 10^{17} ~\mathrm{T_2~molecules}/ \mathrm{cm}^2$ tritium column density both with and without electron catcher operation (setting H, table \ref{tab:before_after_bake_out}). When in operation, the catcher was moved into the flux tube for \SI{5}{s} when the retarding voltage was changed. A reduction of \SI{20}{mcps} was observed when the catcher was used, indicating the presence of a Penning-trap-induced background when the catcher was not used.}
\label{KNM1_rate_w_wo_wiper}
\end{figure}

Figure \ref{KNM1_rate_w_wo_wiper} shows the rate measured when running with tritium in the windowless gaseous tritium source.  Data was taken both with and without periodic electron catcher operation. The column density of the tritium source was about $2.5 \cdot 10^{17} ~\mathrm{T_2~molecules}/ \mathrm{cm}^2$, which corresponds to about $50\%$ of the nominal KATRIN column density and a $\beta$-electron rate of about $5 \cdot 10^9 ~\mathrm{electrons/s}$ entering and being reflected in the pre-spectrometer. For the periodic electron catcher operation, one out of the three electron catchers was inserted into the flux tube for \SI{5}{s} during changes in the retarding voltages, typically every \SI{5}{min}. Figure \ref{KNM1_rate_w_wo_wiper} does not exhibit any indication of discharge-like events, but it also shows that the rate without periodic electron catcher operation is elevated by about \SI{20}{mcps} with respect to the case of periodic electron catcher operation. This difference can be explained by some (beta) electrons producing positive ions before entering the pre-spectrometer. They undergo the above-mentioned sputtering-Rydberg process (see point (e) in section \ref{description}) and create electrons feeding the trap, or stored themselves in the Penning trap. These processes are slow enough for the catchers to be effective.

\section{Conclusion}

We have discussed in this work how the inter-spectrometer Penning trap, which is unavoidable in a two-spectrometer setup, represents a considerable problem for the KATRIN experiment. The trap is a potential limiting factor for the experiment's ability to achieve the required background level. It is also a danger to the spectrometer and detector hardware due to possible discharges when system is operated at pressures higher than the nominal of $\sim 10^{-11}$ mbar.
An effective solution is therefore required. The chosen option of using electron catchers to remove trapped particles was investigated. At pressures in the $10^{-9}$ mbar region two potential modes were tested: a continuously operated electron catcher was able to preserve a quiet environment while a periodically actuated catcher was able to fully quench discharges that occurred between the periodic electron catcher insertions. After the bake-out of the spectrometers, measurements at lower pressures demonstrated a strong dependence of the Penning trap behavior on this parameter. With the finally achieved ultra-high vacuum of $\sim\SI{e-11}{mbar}$, no Penning discharges were observed during a final two-week background measurement as well as during operation with an active tritium source (with $50\%$ of the nominal KATRIN column density), while in the latter, operation of the electron catcher reduced the background rate by about \SI{20}{mcps}. Therefore, the electron catcher setup has proven itself to be a reliable measure against the inter-spectrometer Penning trap effects.

\begin{acknowledgements}
We acknowledge the support of Helmholtz Association, Ministry for Education and Research BMBF (5A17PDA, 05A17PM3, 05A17PX3, 05A17VK2, and 05A17WO3), Helmholtz Alliance for Astroparticle Physics (HAP), Helmholtz Young Investigator Group (VH-NG-1055), and Deutsche Forschungsgemeinschaft DFG (Research Training Groups GRK 1694 and GRK 2149, and Graduate School GSC 1085 - KSETA) in Germany; Ministry of Education, Youth and Sport (CANAM-LM2011019, LTT19005) in the Czech Republic; and the United States Department of Energy through grants DE-FG02-97ER41020, DE-FG02-94ER40818, DE-SC0004036, DE-FG02-97ER41033, DE-FG02-97ER41041, DE-AC02-05CH11231, DE-SC0011091, and DE-SC0019304, and the National Energy Research Scientific Computing Center.
\end{acknowledgements}

\end{document}